\documentclass[twocolumn,english,showpacs,preprintnumbers,amsmath,amssymb,floatfix]{revtex4}
\usepackage[T1]{fontenc}
\usepackage{color}
\usepackage{array}
\usepackage{amstext}
\usepackage{graphicx}
\usepackage{esint}
\usepackage[colorlinks = true,
linkcolor = magenta,
urlcolor  = blue,
citecolor = red,
anchorcolor = blue]{hyperref}

\makeatletter

\@ifundefined{textcolor}{}
{%
	\definecolor{BLACK}{gray}{0}
	\definecolor{WHITE}{gray}{1}
	\definecolor{RED}{rgb}{1,0,0}
	\definecolor{GREEN}{rgb}{0,1,0}
	\definecolor{BLUE}{rgb}{0,0,1}
	\definecolor{CYAN}{cmyk}{1,0,0,0}
	\definecolor{MAGENTA}{cmyk}{0,1,0,0}
	\definecolor{YELLOW}{cmyk}{0,0,1,0}
}

\@ifundefined{definecolor}
{\usepackage{color}}{}
\@ifundefined{definecolor}
{\usepackage{color}}{}
\makeatother

\makeatother

\usepackage{babel}
\begin{document}
	
\title{Charged Higgs production from polarized top-quark decay in the 2HDM considering the general-mass variable-flavor-number scheme}
	
\author{S. Abbaspour$^a$}
	
\author{S. Mohammad Moosavi Nejad$^{a,b}$}
\email{mmoosavi@yazd.ac.ir}

\affiliation{$^{(a)}$Faculty of Physics, Yazd University, P.O. Box
	89195-741, Yazd, Iran\\	
      $^{(b)}$School of Particles and Accelerators,
	Institute for Research in Fundamental Sciences (IPM), P.O.Box
	19395-5531, Tehran, Iran}

	\date{\today}
	
\begin{abstract}

Charged Higgs bosons $H^\pm$ are predicted by some  non-minimal Higgs scenarios, such as models containing Higgs triplets  and two-Higgs-doublet models, so that the experimental observation of these bosons would indicate physics beyond the Standard Model.
In the present work, we introduce a new channel to indirect search for the charged Higgses through the hadronic decay of polarized top quarks  where a  top quark decays into a  charged Higgs $H^+$ and a bottom-flavored hadron $B$ via the hadronization process of the produced bottom quark, $t(\uparrow)\rightarrow H^++b(\to B+jet)$. To obtain the energy spectrum of  produced $B$-hadrons  we present, for the first time, an analytical expression for the ${\cal O}(\alpha_s)$ corrections to the differential decay width of the process $t\rightarrow H^+b$ in the presence of a massive b-quark in  the General-Mass
Variable-Flavor-Number Scheme (GM-VFNS).  We find that  the most reliable predictions for the B-hadron energy spectrum
are made in the GM-VFN  scheme, specifically, when the Type-II 2HDM scenario is concerned.

\end{abstract}

\pacs{14.65.Ha, 13.88.+e, 14.40.Lb, 14.40.Nd}

\maketitle

\section{Introduction}
\label{sec:intro}

There are many reasons, both from experimental observations and  theoretical considerations, to expect physics beyond the Standard Model (BSM), such as the hierarchy problem, neutrino oscillations  and dark matter. Many numerous attempts have been done and are still in progress to build new physics models which can explain these puzzles. Among them, some well-known examples are the Minimal Supersymmetric Standard Model (MSSM) \cite{Inoue:1982pi} and the Two-Higgs-Doublet Models (2HDM) \cite{Lee:1973iz,Gunion}, so the latter is known as the simplest model. In many extensions of the Standard Model (SM) such as the 2HDM, the Higgs sector of the SM is enlarged typically by adding an extra  doublet of complex
Higgs fields. In the 2HDM, after spontaneous symmetry breaking,
the two scalar Higgs doublets $H_1$ and $H_2$ yield three physical
neutral Higgs bosons (h, H, A) and a pair of charged-Higgs bosons $H^\pm$ \cite{Djouadi:2005gj}. \\
The observation of a (singly-)charged-Higgs boson in the current and future runs of the Large Hadron Collider (LHC) would clearly indicate a definitive evidence of new  physics beyond the SM.

From the element $|V_{tb}|\approx 1$ of the Cabibbo-Kobayashi-Maskawa (CKM) quark mixing matrix \cite{Cabibbo:1963yz}, it is found that the top quark  is decaying dominantly through $t\rightarrow bH^+$
in the 2HDM \cite{Lee:1973iz,Gunion}, providing that the top quark mass ($m_t$), bottom quark
mass ($m_b$) and the charged-Higgs boson mass ($m_{H^+}$) satisfy the condition: $m_t>m_b+m_{H^+}$.
In this situation, one may expect measurable effects in the top quark decay width and decay distributions
due to the $H^\pm$-propagator contributions, which are
potentially large in the decay chain $t \rightarrow bH^+\rightarrow b(\tau^+\nu_\tau)$.\\
At the LHC, it is expected to have a cross section $\sigma(pp\rightarrow t\bar{t}X)\approx 1$   (nb) at design energy
$\sqrt{s}=14$ TeV  \cite{Langenfeld:2010zz}. With the LHC design luminosity of $10^{34}$cm$^{-2}$s$^{-1}$ in each of the four experiments, one may expect to have a $t\bar t$ pair  per second so that with this remarkable potential the LHC can be considered as a superlative top factory which allows one to search for the charged-Higgs bosons  in the subsequent decay
products of the top pairs $t\bar t\rightarrow H^\pm W^\mp b\bar b$ and $t\bar t\rightarrow H^\pm H^\mp b\bar b$. New results of a search on the charged-Higgs bosons in the proton-proton collision at a center of mass
energy of $\sqrt{s}=8$ TeV  are reported by the CMS \cite{CMS:2014cdp} and ATLAS \cite{TheATLAScollaboration:2013wia} Collaborations at the LHC, and we shall discuss on these results and the restrictions imposed  over the MSSM $m_{H^+}-\tan\beta$ parameter space in Section \ref{sec:three}. In our numerical analysis we restrict ourselves to the unexcluded regions of this parameter space.

As was mentioned, due to  $|V_{tb}|\approx 1$ of the CKM matrix the top quark almost exclusively decays to b-quark via $t\to bH^+$, in the 2HDM. On the other hand, b-quarks hadronize, via $b\to B+X$, before they decay so that the decay process $t\to BH^++X$ is of prime importance and it is an urgent task to predict its partial decay width  as realistically and reliably as possible. Therefore, at the LHC of particular interest  would be the distribution in the scaled B-hadron energy ($x_B$) in the top quark rest frame so that  the study of this distribution is proposed as a new channel to indirect search for the charged Higgs bosons. To study the scaled energy spectrum of B-hadrons we determine the quantity $d\Gamma(t\to BH^++X)/dx_B$.\\
According to the factorization  theorem of the QCD-improved parton model \cite{collins}, the B-hadron energy
distribution  can be determined by  the convolution of the partonic differential decay width $(d\Gamma/dx_b)$ of the subprocess $t\to bH^+$, with
the nonperturbative fragmentation function (FF) $D_b^B$, as
\begin{eqnarray}\label{azi}
\frac{d\Gamma}{dx_B}=\frac{d\Gamma}{dx_b}(\mu_R,\mu_F)\otimes D_b^B(\frac{x_B}{x_b}, \mu_F),
\end{eqnarray}
where,  $x_b$ stands for the scaled-energy fraction of the bottom quark and the $D_b^B$-FF describes the splitting process $b\to B+X$ in which $X$ refers to the unobserved final state particles.
In (\ref{azi}),  $\mu_F$ and $\mu_R$ are the factorization and renormalization scales, respectively, and the integral convolution is defined as $(f\otimes g)(x)=\int_x^1 dy/y f(y)g(x/y)$.

In \cite{MoosaviNejad:2011yp}, we studied  the energy spectrum of the bottom-flavored mesons through unpolarized top quark decays in the 2HDM at next-to-leading order (NLO) of the QCD radiative corrections.
In \cite{MoosaviNejad:2016jpc}, we studied the spin-dependent energy distribution of B-mesons produced through polarized top  decays   in the massless scheme or Zero-Mass Variable-Flavor-Number scheme (ZM-VFNs)  \cite{Binnewies:1998vm} where the zero mass parton approximation is also applied to the bottom quark.  This massless approximation simplifies the evaluation of the NLO QCD corrections largely but at the price of losing the accuracy of analysis. For example, by this approximation the results for the energy spectrum of B-mesons are independent of the model selected.

In the present work, we impose the effects of bottom quark mass
on the spin-dependent energy spectrum of B-mesons  employing the General-Mass Variable-Flavor-Number (GM-VFN) scheme in which we preserve the b-quark mass from the beginning. Our calculations are done in a special helicity frame where the polarization direction of the top quark is evaluated with
respect to the b-quark three-momentum $\vec{p}_b$.   As will be shown,
the results are different in two variants of the 2HDM and it is found that the NLO corrections with $m_b\neq 0$
to be significant, specifically, when the type-II 2HDM is considered.

In the SM, due to $|V_{tb}|\approx 1$ the top quark  decays dominantly through the decay mode $t\rightarrow bW^+$. In  \cite{Kniehl:2012mn,Nejad:2015pca,Nejad:2016epx,Nejad:2013fba,Nejad:2014sla}, we investigated the  energy distribution of B-mesons produced in polarized and unpolarized top quark decays in the SM. For the unpolarized top decays, the total distribution of the B-hadron energy  is obtained by the summation of  two contributions due to the decay modes $t\rightarrow bH^+$ (in the 2HDM) and $t\rightarrow bW^+$ (in the SM), i.e. $d\Gamma_{\mathrm{tot}}/dx_B=d\Gamma^{\mathrm{SM}}(t\to  BW^+)/dx_B+d\Gamma^{\mathrm{BSM}}(t\to BH^+)/dx_B$. 
The same result is valid for the polarized top quark decay as long as the spin direction of the polarized top quark is evaluated relative to the b-quark three-momentum, as in our present work. Thus, at the LHC any deviation of the B-meson energy spectrum from the SM predictions can be considered as a signal for the existence of charged Higgs.  Although,
the SM contribution  is normally larger than the one coming from the 2HDM (this comparison is studied in Ref.~\cite{MoosaviNejad:2011yp}), but there is always a clear separation between the decay channels $t\to bW^+$ and $t\to bH^+$ in both the $t\bar{t}X$ pair production and the $t/\bar{t}X$ single top production at the LHC, this point is mentioned in \cite{Ali:2011qf}.

As a last point; for the decay chain $t \rightarrow bH^+\rightarrow b(\tau^+\nu_\tau)$ one has $\Gamma(t \to b\tau^+\nu_\tau)=\Gamma(t\to bH^+)\times BR(H^+\to \tau^+\nu_\tau)$ in the narrow width approximation for the Higgs boson. For the numerical values applied in this work, the branching ratio $BR(H^+\to \tau^+\nu_\tau)=1$ \cite{Ali:2009sm}, to a very high accuracy. Therefore, the results presented in this work are also valid for the decay chain $t(\uparrow) \to B\tau^+\nu_\tau+X$.

This paper is organized as follows.
In Sec.~\ref{sec:angular}, we introduce the general angular structure of the differential decay width in a specific helicity frame.   In Sec.~\ref{sec:one}, we present our analytical results of the ${\cal O}(\alpha_s)$ QCD corrections to the tree-level rate of $t(\uparrow)\rightarrow bH^+$ in the fixed flavor number scheme.
In Sec.~\ref{sec:two}, in a detailed discussion we describe the GM-VFN scheme by introducing the perturbative FF $b\to b$. In Sec.~\ref{sec:three}, our hadron level results  in the GM-VFN scheme will be presented.
In Sec.~\ref{sec:four},  our conclusions are summarized.

\section{Angular structure of differential decay rate}
\label{sec:angular}

Here, we concentrate on the decay process $t(\uparrow)\rightarrow bH^+$ in the general 2HDM, where  $H_1$ and $H_2$ are the
doublets whose vacuum expectation values give masses to the down and up type quarks, respectively, and
a linear combination of the charged components of $H_1$ and $H_2$ also gives the physical charged Higgs $H^+$ ($H^+=\cos\beta H_2^+-\sin\beta H_1^+$). The parameter $\beta$ is defined in the following.

Basically, the dynamics of the current-induced $t\rightarrow b$ transition is embodied in the hadronic tensor 
$H^{\mu\nu}\propto \sum_{X_b}\left\langle t(p_t, s_t)|J^{\mu\dagger}|X_b\right\rangle\left\langle X_b|J^\nu|t(p_t, s_t)\right\rangle$, where $s_t$ denotes the spin of the top quark.
At the NLO QCD radiative corrections, only two types of intermediate states are contributed; $|X_b>=|b>$ for the Born level
term and ${\cal O}(\alpha_s)$ one-loop contributions and $|X_b>=|b+g>$ for the ${\cal O}(\alpha_s)$ tree graph contribution.  
In the SM, where one has $t\to bW^+$, the weak current is given by $J^\mu=(J_\mu^V-J_\mu^A)\propto \bar{\psi}_b\gamma^\mu(1-\gamma_5)\bar{\psi}_t $ while in the 2HDM, the current is given by $J^{\mu}\propto \bar{\psi}_b(a+b\gamma_5)\bar{\psi}_t$.

Generally, in models with two Higgs doublets and generic coupling to all the quarks, it is difficult to avoid tree level flavor changing neutral currents, therefore, we limit ourselves to models that naturally stop these problems by restricting the Higgs coupling. As may be found in \cite{GHK}, the first possibility  is to have  the doublet $H_1$ coupling  to all bosons and the doublet $H_2$ coupling to all the quarks (called model I in the following). This model leads to the coupling factors as
\begin{eqnarray}\label{yak}
a&=&\frac{g_W}{2\sqrt{2}m_W}V_{tb}(m_t-m_b)\cot\beta,\nonumber\\
b&=&\frac{g_W}{2\sqrt{2}m_W}V_{tb}(m_t+m_b)\cot\beta,
\end{eqnarray}
where, $\tan\beta=\textbf{v}_2/\textbf{v}_1$ is  the ratio of the vacuum expectation values of the two electrically neutral components of the two Higgs doublets and the weak coupling factor $g_W$ is related to the Fermi's coupling constant by $g_W^2=4\sqrt{2}m_W^2G_F$.\\
In the second possibility (called model II), the doublet $H_2$ couples to the right-chiral up-type quarks ($u_R, c_R, t_R$) while the $H_1$ couples to  the right-chiral down-type quarks. In this model the coupling factors read
\begin{eqnarray}\label{do}
a&=&\frac{g_W}{2\sqrt{2}m_W}V_{tb}(m_t \cot\beta+m_b\tan\beta),\nonumber\\
b&=&\frac{g_W}{2\sqrt{2}m_W}V_{tb}(m_t \cot\beta-m_b\tan\beta).
\end{eqnarray}
These models are mostly known as Type-I and Type-II 2HDM scenarios. The type-II is the Higgs sector of the Minimal Supersymmetric Standard Model (MSSM) up to SUSY corrections \cite{Inoue:1982pi}. Two other models (models III and IV) are also possible which are explained in our previous work \cite{MoosaviNejad:2016jpc} in detail. See also \cite{Barger:1989fj}. 
Here, we just mention that the analytical results presented for the partonic process $t(\uparrow)\to bH^+$ are the same both in models I and IV and also in models II and III.\\
In the rest frame of a polarized top quark decaying into a b-quark and a Higgs boson (and a gluon at NLO), the final-state particles $(b, H^+, g)$
define an event frame. Relative to this event plane, the polarization direction of  top quark can be defined. 
In this work, we analyze the decay mode $t(\uparrow)\to bH^+(+g)$ in the rest frame of the polarized top quark where the three-momentum of the bottom quark points into the direction of the positive $\hat{z}$-axis and the polar angle $\theta_P$ is defined as the angle between the polarization vector of top quark and the $\hat{z}$-axis, see Fig.~1 of Ref.~\cite{MoosaviNejad:2016jpc}.

The general angular distribution of the differential decay width $d\tilde\Gamma/dx_b$ of a polarized top quark is given  by the following
form  to clarify the correlation between the polarization of the top quark and its decay products  
\begin{eqnarray}\label{form}
\frac{d^2\tilde\Gamma}{dx_bd\cos\theta_P}=\frac{1}{2}\bigg\{\frac{d\tilde\Gamma^{\textbf{unpol}}}{dx_b}+P\frac{d\tilde\Gamma^{\textbf{pol}}}{dx_b}\cos\theta_P\bigg\},
\end{eqnarray}
where the polar angle $\theta_P$ shows the spin orientation of the top quark relative to the event plane and $P( 0\leq P\leq 1)$ is the
magnitude of the top quark polarization. In the notation above, the first and second terms in the curly bracket  correspond to the unpolarized and polarized differential decay rates, respectively. 
As usual, we define the scaled-energy fraction of the bottom quark  as
\begin{eqnarray}\label{vari}
x_b=\frac{E_b}{E_b^{max}}=\frac{2m_tE_b}{m_t^2+m_b^2-m_{H^+}^2}.
\end{eqnarray}
The ${\cal O}(\alpha_s)$ radiative corrections to the unpolarized differential width $d\tilde\Gamma^{\textbf{unpol}}/dx_b$ have been 
studied in \cite{MoosaviNejad:2011yp}, extensively, and the NLO QCD corrections to the polarized partial rate $d\hat\Gamma^{\textbf{pol}}/dx_b$ in the ZM-VFN scheme (with $m_b=0$) are studied in \cite{MoosaviNejad:2016jpc}.
In the present work, we compute the NLO QCD radiative corrections to the polarized partial rate $d\tilde\Gamma^{\textbf{pol}}/dx_b$ in the GM-VFN scheme where $m_b\neq 0$ is considered from the beginning. These analytical results are new and presented for the first time.

\section{Parton level results}
\label{sec:one}

\subsection{Born term results}

It is straightforward to compute the Born term contribution to the partial decay rate of the polarized top quark in the 2HDM in the presence of the b-quark mass.
The Born term amplitude for the process $t(\uparrow)\rightarrow bH^+$  can be parameterized
as \textit{$M_0=\bar{u}_b(a+b\gamma_5)u_t$}, so for the  squared amplitude one has: $|M_0|^2=2(p_b\cdot p_t)(a^2+b^2)+
2(a^2-b^2)m_bm_t+4ab m_t(p_b\cdot s_t)$, where we replaced $\sum_{s_t}u(p_t, s_t)\bar{u}(p_t, s_t)=(\displaystyle{\not}{p}_t+m_t)$ in the unpolarized Dirac string by $u(p_t, s_t)\bar{u}(p_t, s_t)=(1-\gamma_5 \displaystyle{\not}{s}_t )(\displaystyle{\not}{p}_t+m_t)/2$ in the polarized state.
Then, the polarized tree-level decay width reads
\begin{eqnarray}\label{gammatree}
\tilde\Gamma_{0P}&=&\tilde\Gamma_{Born}^{\textbf{pol}}=\frac{1}{16\pi m_t}\lambda^{1/2}(1,\frac{m_b^2}{m_t^2},
\frac{m_{H^+}^2}{m_t^2})\times\nonumber\\
&&\bigg(2abm_t^2\lambda^{1/2}(1,\frac{m_b^2}{m_t^2},
\frac{m_{H^+}^2}{m_t^2})\bigg)=\frac{m_t\lambda}{8\pi}(ab),
\end{eqnarray}
where $\lambda=\lambda(x,y,z)=(x-y-z)^2-4y z$ is the  K\"all\'en function and the factor $\sqrt{\lambda}/(16\pi m_t)=PS_2$ is the two-body phase space factor.
This result is in complete agreement with  Refs.~\cite{kadeer,Liud}.
The unpolarized Born-level rate can be found in our previous work \cite{MoosaviNejad:2011yp}.
Considering Eqs.~(\ref{yak}) and (\ref{do}), in (\ref{gammatree}) for the product of two coupling factors in the model I (type-I 2HDM scenario), one has
\begin{eqnarray}\label{haselmodel1}
ab=\frac{G_F}{\sqrt{2}} |V_{tb}|^2 m_t^2(1-\frac{m_b^2}{m_t^2})\cot^2\beta,
\end{eqnarray}
and for the model II (type-II 2HDM scenario),
\begin{eqnarray}\label{haselmodel2}
ab=\frac{G_F}{\sqrt{2}} |V_{tb}|^2 m_t^2(1 -\frac{m_b^2}{m_t^2} \tan^4\beta)\cot^2\beta.
\end{eqnarray}
Since $m_b\ll m_t$, the bottom quark mass can always be safely neglected in the model I, while in the model II, the second term in (\ref{haselmodel2}) can become comparable to the first term when $\tan\beta$ becomes large, then one can not naively set $m_b=0$ in all expressions. 
For instance, if one takes $m_b=4.78$~GeV, $m_t=172.98$~GeV, $m_{H^+}=155$~GeV and $\tan\beta=5$ thus the second term in the parenthesis  (\ref{haselmodel2})  can become as large as ${\cal O}(48\%)$ and this order will be larger when $\tan\beta$ is increased.
Therefore, the results in the type-II 2HDM scenario depend on the b-quark mass extremely, unless the low values of  $\tan\beta(\tan\beta\leq 1)$ are  applied, however, these small values of  $\tan\beta$ are now excluded by the CMS and ATLAS experiments at the LHC. 
In \cite{MoosaviNejad:2016jpc}, we adopted  the ZM-VFN scheme (with $m_b= 0$) which is not suitable for the Type-II 2HDM scenario. There, we pointed out that our phenomenological predictions are restricted to the Type-I 2HDM. In the present work we retain the b-quark mass and extend our results to both models and shall compare them. 

In the following,  in a detailed discussion we calculate the ${\cal O}(\alpha_s)$ QCD corrections to the Born-level width and
 present an analytical expression for $d\Gamma(t(\uparrow)\rightarrow BH^++X)/dx_B$ at
NLO in the GM-VFN scheme.

\subsection{Virtual gluon corrections including counterterms and one-loop vertex correction}\label{virtual}

Basically, the one-loop virtual corrections  to the $tbH^+$-vertex consist of both the infrared (IR) and the
ultraviolet (UV) singularities. The UV-divergences appear when the integration
region of the virtual gluon momentum goes to infinity and the
IR-divergences arise from the soft-gluon singularities.
Here, we adopt
the on-shell mass-renormalization scheme and all singularities are regularized by dimensional
regularization in $D=4-2\epsilon$ space-time dimensions where $\epsilon\ll 1$.
Using the dimensional regularization technique one obtains the well-defined integrals which are finite  while all singularities are summarized in the $\epsilon$. This is done by 
the replacement:  $\int d^4 p_g/(2\pi)^4\rightarrow \mu^{4-D}\int d^D p_g/(2\pi)^D$ in the  one-loop integrals, where
 $\mu$ is an arbitrary reference
mass which will be removed after summing all corrections up.\\
For simplicity, we introduce  the following abbreviations:
\begin{eqnarray}
	S&=&\frac{1}{2}(1+R-y),
	\nonumber\\
	\beta &=&\frac{\sqrt{R}}{S},
	\nonumber\\
	Q&=&S\sqrt{1-\beta^2},
	\\
	\Phi(x_b)&=&S \big[\sqrt{x_b^2-\beta^2}-\ln\frac{\beta}{x_b-\sqrt{x_b^2-\beta^2}}\big],
	\nonumber\\
	T&=&S(1-x_b)\sqrt{x_b^2-\beta^2}+x_b \Phi(x_b),\nonumber
\end{eqnarray}
where the scaled masses  $y=m_{H^+}^2/m_t^2$ and $R=m_b^2/m_t^2$ are defined. Choosing
these notations,  the tree-level total width (\ref{gammatree}) is simplified as 
\begin{eqnarray}\label{newgammatree}
\tilde\Gamma_{0P}=\frac{m_t Q^2}{2\pi}(ab).
\end{eqnarray}
Also, the normalized energy
fraction $x_b$ (\ref{vari}) is given by 
\begin{eqnarray}\label{variabblee}
x_b=\frac{E_b}{m_t S}.
\end{eqnarray}

Taking the above notations, the contribution of virtual corrections into the  decay width reads
\begin{eqnarray}
	\frac{d\tilde\Gamma^{\textbf{vir,pol}}}{dx_b}=\frac{Q}{8\pi m_t}
	|M^{\textbf{vir}}|^2\delta(1-x_b),
\end{eqnarray}
where,
$|M^{\textbf{vir}}|^2=\sum_{Spin}(M_0^{\dagger} M_{loop}+M_{loop}^{\dagger} M_0)$ and the factor $PS_2=Q/(8\pi m_t)$ is a two-body phase space factor, as in (\ref{gammatree}).
The renormalized amplitude  is now written as
$M_{loop}=\bar{u}_b(\Lambda_{ct}+\Lambda_l)u_t$,
where   $\Lambda_l$
arises from the one-loop vertex correction and $\Lambda_{ct}$ stands for the counter term.
The counter term of the vertex consists of the mass and the wave function renormalizations of both the top and bottom quarks \cite{kadeer,Czarnecki,Liud}, as
\begin{eqnarray}
	\Lambda_{ct}&=&(a+b) \bigg(\frac{\delta Z_b}{2}+\frac{\delta Z_t}{2}-
	\frac{\delta m_t}{m_t}\bigg)\frac{1+\gamma_5}{2}
	\nonumber\\
	&&+(a-b) \bigg(\frac{\delta Z_b}{2}+\frac{\delta Z_t}{2}-\frac{\delta m_b}{m_b}\bigg)\frac{1-\gamma_5}{2},
\end{eqnarray}
where, the  wave function and the mass renormalization constants  are expressed as 
\begin{eqnarray}\label{mass}
	\delta Z_q &=& -\frac{\alpha_s(\mu_R)}{4\pi}C_F\big[\frac{1}{\epsilon_{UV}}+\frac{2}{\epsilon_{IR}}
	-3\gamma_E+3\ln\frac{4\pi\mu_F^2}{m_q^2}+4\big],
	\nonumber\\
	\frac{\delta m_q}{m_q}&=&\frac{\alpha_s(\mu_R)}{4\pi}C_F\big[\frac{3}{\epsilon_{UV}}-3\gamma_E+
	3\ln\frac{4\pi\mu_F^2}{m_q^2}+4\big].
\end{eqnarray}
In the equation above,  $m_q (q=b,t)$ is the mass of the relevant quark,  $\gamma_E=0.5772\cdots$ is the Euler constant
and  $C_F=(N_c^2-1)/(2N_c)=4/3$ for $N_c=3$ quark colors.
Also, $\epsilon_{IR}$ and $\epsilon_{UV}$ represent infrared and
ultraviolet singularities.

The real part of the one-loop vertex correction is given by
\begin{eqnarray}
	\Lambda_l=\frac{\alpha_s m_t^2}{\pi Q}C_F (2ab)G(m_b^2,m_t^2,m_{H^+}^2),
\end{eqnarray}
with
\begin{eqnarray}
	G&=&2Q^2-4m_t^2 S Q^2 C_0(m_b^2,m_t^2,m_{H^+}^2,m_b^2,0,m_t^2)
	\nonumber\\
	&&+(R+RS-2S^2)B_0(m_b^2,0,m_b^2)\nonumber\\
	&&+(2R-S-RS)B_0(m_{H^+}^2,m_b^2,m_t^2)
	\nonumber\\
	&&+(R+S-2S^2)B_0(m_t^2,0,m_t^2),
\end{eqnarray}
where, $B_0$ and $C_0$ functions are the Passarino-Veltman 2-point and 3-point integrals, respectively. The analytical form of these integrals can  be found in Ref.~\cite{Dittmaier:2003bc}.\\
All the  ultraviolet divergences   shall be canceled after summing all virtual corrections up but the infrared singularities ($\epsilon_{IR}$)
are remaining which are now labeled by $\epsilon$.
Putting everything together, for the virtual differential decay width  normalized to the Born-level total rate (\ref{newgammatree}) one has
\begin{eqnarray}\label{virt}
	\frac{1}{\tilde\Gamma_{0P}}\frac{d\tilde\Gamma^{\textbf{vir,pol}}}{dx_b}&&=
	-\frac{\alpha_s(\mu_R)}{4\pi Q}C_F
	\delta(1-x_b)\bigg\{\frac{Q}{2}\ln R\bigg[\frac{8S}{y}
	\nonumber\\
	&&\hspace{-1cm}-\frac{8}{y}+\frac{3 (a^2+b^2)}{ab}-\frac{2S}{Q}\ln\sqrt{R}\bigg]+2S\ln^2(S+Q)
	\nonumber\\
	&&\hspace{-1cm} +S\ln\frac{S+Q}{S-Q}\bigg(-\ln\frac{S-Q}{R\sqrt{R}}-2\frac{RS+S-2R}{Sy}\bigg)
	\nonumber\\
	&&\hspace{-1cm}-2\bigg[S\ln\frac{S+Q}{S-Q}-2Q\bigg]\bigg(\ln\frac{4\pi \mu_F^2}{m_t^2}-
	\gamma_E+\frac{1}{\epsilon}\bigg)
	\nonumber\\
	&&\hspace{-1cm}+4S\bigg[Li_2(S-Q)-Li_2(S+Q)+Li_2(\frac{2Q}{S+Q})\nonumber\\
	&&\hspace{-1cm}+\ln y\ln(S-Q)-\ln R\ln(1-S-Q)+\frac{Q}{S}\bigg]\bigg\}.
	\nonumber\\
\end{eqnarray}
This result after integration over $x_b$ is in complete agreement with Ref.~\cite{kadeer}.

\subsection{Real gluon radiative corrections}\label{real}

If we denote the polarization vector of the real gluon  by $\epsilon(p_g, \lambda)$, the  ${\cal O}(\alpha_s)$ real gluon emission (tree-graph) amplitude reads
\begin{eqnarray}
	M_{tree}&=&g_s\frac{\lambda^a}{2}\bar{u}_b\big\{\frac{2p_t^\sigma-
		\displaystyle{\not}p_g \gamma^\sigma}{2p_t \cdot p_g}-\frac{2p_b^\sigma+\gamma^\sigma\displaystyle{\not}p_g}
	{2p_b\cdot p_g}\big\}\nonumber\\
	&&\times(a\textbf{1}+b\gamma_5) u_t\epsilon_\sigma^{\star}(p_g,\lambda).
\end{eqnarray}
In the ZM-VFN scheme, the IR-divergences arise from the soft- and collinear gluon emissions while in the GM-VFN scheme there are  no collinear divergences  and all IR-singularities arise from the soft real gluon emission.
As before, to regulate the IR-divergences we work in $D$-dimensions so that the contribution of real gluon emission into the polarized differential decay rate is given by
\begin{eqnarray}\label{ree}
	d\tilde\Gamma^{\textbf{real,pol}}=\frac{1}{2m_t}\frac{\mu_F^{2(4-D)}}{(2\pi)^{2D-3}}
	dPS_3(p_t, p_H, p_b, p_g)|M^{\textbf{tree}}|^2,\nonumber\\
\end{eqnarray}
where $dPS_3$ is the three-body phase space
\begin{eqnarray}
	dPS_3=\prod_{i=b,g,H}\frac{d^{D-1}\textbf{p}_i}{2E_i}\delta^D(p_t-\sum_{i=b,g,H}p_i).
\end{eqnarray}
To compute the real differential decay rate $d\tilde\Gamma^{\textbf{real}}/dx_b$, 
in (\ref{ree}) the momentum of $b$-quark is fixed  and  over the energy
of the gluon is integrated. The energy of gluon ranges from  $E_g^{min}=F(1-S x_b-S\sqrt{x_b^2-\beta^2})$ to 
$E_g^{max}=F(1-S x_b+S\sqrt{x_b^2-\beta^2})$, where $F=m_t S (1-x_b)/(1+R-2S x_b)$.\\
To achieve the correct finite terms in the rate $1/\tilde{\Gamma}_{0P}\times d\tilde{\Gamma}^{\textbf{real,pol}}/dx_b$, 
the Born width ${\tilde\Gamma}_{0P}$ (\ref{gammatree}) 
must be evaluated in
the dimensional regularization at ${\cal O}(\epsilon)$, i.e.
$\tilde{\Gamma}_{0P}\rightarrow \tilde{\Gamma}_{0P}\{1-\epsilon
\big[2\ln Q+\gamma_E-\ln(4\pi\mu_F^2/m_t^2)\big]\}$.
When integrating  over the phase space, terms of the
form $(1-x_b)^{-1-2\epsilon}$ arise which are due to  the radiation of a soft gluon in  top decay at NLO.
Note, the  $E_g\rightarrow 0$ limit corresponds to the limit $x_b\rightarrow 1$.
Thus for a massive bottom quark, where $x_{b,min}=\beta$, we apply the following expansion \cite{Corcella:1} 
\begin{eqnarray}
	\frac{(x_b-\beta)^{2\epsilon}}{(1-x_b)^{1+2\epsilon}}=-\frac{1}{2\epsilon}\delta(1-x_b)+
	\frac{1}{(1-x_b)_+}+{\cal O}(\epsilon),
\end{eqnarray}
where the plus distribution  is defined as usual.

Finally, the contribution of real gluon emission reads
\begin{eqnarray}\label{reall}
	&&\frac{1}{\tilde\Gamma_{0P}}\frac{d\tilde\Gamma^{\textbf{real,pol}}}{dx_b}=
	\frac{C_F\alpha_s(\mu_R)}{2\pi Q}\Bigg\{
	\nonumber\\
	&&\hspace{+0.8cm}\delta(1-x_b)\bigg[-4Q\ln\frac{2S}{\sqrt{y}}+(R-1)\ln\frac{1-S-Q}{1+Q-S}
	\nonumber\\
	&&\hspace{+0.8cm}-2S\big[Li_2(S-Q)-Li_2(S+Q)+Li_2(\frac{2Q}{S+Q})\big]+
	\nonumber\\
	&&\bigg(S\ln\frac{Q+S}{S-Q}-2Q\bigg)\bigg(\gamma_E-\frac{1}{\epsilon}-\ln\frac{4\pi\mu_F^2}{m_t^2}+2\ln(1-\beta)\bigg)
	\nonumber\\
	&&\hspace{+0.6cm} -S\ln\frac{S+Q}{S-Q}\bigg(-1+\frac{R}{S}-2\ln(2S)+\frac{1}{2}\ln\frac{S+Q}{S-Q}\bigg)\bigg]
	\nonumber\\
	&&\hspace{0.8cm}-\frac{2S(1+x_b)T}{Q\sqrt{x_b^2-\beta^2}}-\frac{4S T \sqrt{x_b^2-\beta^2}}{Q(1-x_b)_+}\Bigg\}.
\end{eqnarray}
To obtain an analytic result for the polarized partial decay rate, by summing the tree level, the virtual and the real contributions, one has
 \begin{eqnarray}\label{first}
	\frac{1}{\tilde\Gamma_{0P}}\frac{d\tilde\Gamma^{\textbf{pol}}}{dx_b}&&=\delta(1-x_b)+
	\frac{C_F\alpha_s(\mu_R)}{2\pi Q}\Bigg\{
	\nonumber\\
	&&\delta(1-x_b)\bigg[-2Q-4Q\ln\frac{2S(1-\beta)}{\sqrt{y}}+
	\nonumber\\
	&&4S\big[Li_2(S+Q)-Li_2(S-Q)-Li_2(\frac{2Q}{S+Q})\big]
	\nonumber\\
	&&+(R-1)\ln\frac{1-S-Q}{1+Q-S}-\ln R\bigg(\frac{3(a^2+b^2)Q}{4 ab}
	\nonumber\\
	&&+\frac{2Q(S-1)}{y}+S\ln\frac{1+Q-S}{1-S-Q}\bigg)-
	\nonumber\\
	&& 4S\ln\frac{S-Q}{\sqrt{R}}\bigg(\ln(2S(1-\beta))+\frac{1}{2}\ln\frac{y}{R}+\nonumber\\
	&&\ln\frac{S-Q}{\sqrt{R}}+\frac{S(1+y)+R(S-y-2)}{2Sy}\bigg)\bigg]
	\nonumber\\
	&&-T\bigg(\frac{2S(1+x_b)}{Q\sqrt{x_b^2-\beta^2}}+\frac{4S \sqrt{x_b^2-\beta^2}}{Q(1-x_b)_+}\bigg)\Bigg\}.
\end{eqnarray}
As is seen,  all  IR-singularities are canceled and the final result is free of singularities. 

In Ref.~\cite{kadeer}, authors considered a specific helicity coordinate system where the polarization vector of the top quark was evaluated relative to the Higgs boson three-momentum. In \cite{MoosaviNejad:2016aad}, we applied the same frame and obtained the polarized differential decay width $d\hat{\Gamma}^{\textbf{pol}}/dx_b$ at the parton-level in the ZM-VFN scheme to obtain the energy spectrum of B-hadrons (according to Eq.~(\ref{azi})). There, we showed that our analytical result for the parton-level differential decay rate $d\hat{\Gamma}^{\textbf{pol}}/dx_b$ is in agreement with \cite{kadeer} after integration over $x_b(0\leq x_b \leq 1)$.  In this work, following our previous work \cite{MoosaviNejad:2016jpc},  we considered a new helicity frame where the polarization vector of the top quark is evaluated relative to the b-quark three-momentum. Applying the same techniques,  our results for the Born rate (\ref{newgammatree}) and virtual corrections (\ref{virt}) are the same in both helicity frames but the real corrections (\ref{reall}) and, in conclusion, the NLO differential decay width (\ref{first}) are different and completely new. In next section we present our reason for correctness of the obtained result.

\section{General-Mass Variable-flavor-number scheme}
\label{sec:two}

In this work, our main purpose is to evaluate the scaled-energy distribution of the B-hadron produced in
the inclusive process $t(\uparrow)\rightarrow BH^++X$ in the 2HDM.
Therefore, we calculate the NLO decay width  of the corresponding process
differential in $x_B$ ($d\Gamma^{\textbf{pol}}/dx_B$)  in the GM-VFN scheme, where $x_B=E_B/(m_t S)$ is
the scaled-energy fraction of the B-hadron (as for $x_b$ in (\ref{variabblee})).
In the top quark rest frame applied in our work, the B-hadron has the energy $E_B=p_t\cdot p_B/m_t $, where
$m_B\le E_B\le [m_t^2+m_B^2-m^2_{H^+}]/(2m_t) $.\\
Considering the factorization  theorem (\ref{azi}), the B-hadron energy
spectrum  can be obtained by  the convolution of the parton-level spectrum (\ref{first}) with
the nonperturbative fragmentation function (FF) $D_b^B(z, \mu_F)$. We will discuss about the FFs needed, later.

We now explain how the quantity
$d\Gamma^{\textbf{pol}}(\mu_R, \mu_F) /dx_b$ will have to be evaluated in the GM-VFN scheme.
In Sec.~\ref{sec:one}, we employed the Fixed-Flavor-Number (FFN) scheme which
contains of the full $m_b$ dependence.  In the FFN scheme,
the large logarithmic singularities of the form $(\alpha_s/\pi)\ln(m_t^2/m_b^2)$  spoil the
convergence of the perturbative expansion when $m_b/m_t\rightarrow 0$. The massive or GM-VFN scheme is devised to resum these large logarithms  and to retain the whole
nonlogarithmic $m_b$ dependence at the same time and this is achieved by introducing
appropriate subtraction terms in the NLO FFN expression for $d\tilde\Gamma^{\textbf{pol}}/dx_b$. In this case, the NLO ZM-VFN result is  exactly recovered in the limit $m_b/m_t\rightarrow 0$.
In the GM-VFN scheme, the subtraction term is  constructed as
\begin{eqnarray}\label{mohsen}
	\frac{1}{\Gamma_{0P}}\frac{d\Gamma^{\textbf{pol}}_{\textbf{Sub}}}{dx_b}=\lim_{m_b\rightarrow 0}
	\frac{1}{\tilde\Gamma_{0P}}\frac{d\tilde\Gamma_{\textbf{FFN}}^{\textbf{pol}}}{dx_b}-
	\frac{1}{\hat\Gamma_{0P}}\frac{d\hat\Gamma_{\textbf{ZM}}^{\textbf{pol}}}{dx_b},
\end{eqnarray}
where $1/\hat\Gamma_{0P}\times d\hat\Gamma_{\textbf{ZM}}^{\textbf{pol}}/dx_b$ is the partial decay rate computed in the ZM-VFN scheme \cite{MoosaviNejad:2016jpc}, in which all information
on the $m_b$-dependence of $d\hat\Gamma^{\textbf{pol}}/dx_b$  is wasted.\\
In conclusion,  the GM-VFN result is obtained by subtracting the subtraction term from the FFN one \cite{Kniehl:2,Kniehl:3}, 
\begin{eqnarray}
	\frac{1}{\Gamma_{0P}}\frac{d\Gamma^{\textbf{pol}}_{\textbf{GM}}}{dx_b}=
	\frac{1}{\tilde\Gamma_{0P}}\frac{d\tilde\Gamma_{\textbf{FFN}}^{\textbf{pol}}}{dx_b}-
	\frac{1}{\Gamma_{0P}}\frac{d\Gamma_{\textbf{Sub}}^{\textbf{pol}}}{dx_b}.
\end{eqnarray}
Taking the limit $m_b\rightarrow 0$ in Eq.~(\ref{first}), one obtains  the following subtraction term 
\begin{eqnarray}\label{bff}
	\frac{1}{\Gamma_{0P}}\frac{d\Gamma_{\textbf{Sub}}^{\textbf{pol}}}{dx_b}&=&\frac{\alpha_s(\mu_R)}{2\pi}C_F\times
	\nonumber\\
	&&\hspace{-1cm}\bigg\{\frac{1+x_b^2}{1-x_b}\bigg[\ln\frac{\mu_F^2}{m_b^2}-2\ln(1-x_b)-1\bigg]\bigg\}_+.
	\end{eqnarray}
This result coincides with the perturbative FF of the
transition $b\rightarrow b$ \cite{Mele:1990cw} and is in consistency  with the  Collin's factorization theorem \cite{collins}, which guarantees that the subtraction terms are universal. Thus, the result obtained in  (\ref{bff}) ensures the correctness of our result shown in  (\ref{first}).

\section{Numerical analysis}
\label{sec:three}

In the MSSM, the mass of charged Higgses is restricted  by $m_{H^\pm}>m_{W^{\pm}}$ at tree-level \cite{Nakamura:2010zzi}, however, this restriction is not valid  for some regions of parameter space after including radiative corrections.
In the MSSM, $m_{H^\pm}$ is strongly correlated with the mass of other Higgs bosons.
In Ref.~\cite{Ali:2009sm} is mentioned that 
a charged Higgs with a mass  range $80~ GeV\leq m_{H^\pm}\leq 160~ GeV$ is a logical possibility
and its effects should be searched for in the decay chain $t\rightarrow  bH^+\rightarrow B\tau^+\nu_\tau+X$.\\
On the other side, the  last results of a search for evidence of a light charged Higgs boson ($m_H<m_t$)  in $19.5-19.7 fb^{-1}$ of proton-proton collision data recorded at $\sqrt{s}=8$~TeV are reported by the CMS \cite{CMS:2014cdp} and the ATLAS \cite{TheATLAScollaboration:2013wia}  collaborations, using the $\tau+jets$ channel with a hadronically decaying $\tau$ lepton in the final state. 
According to Fig.~7 of  Ref.~\cite{TheATLAScollaboration:2013wia}, the large region in the MSSM $m_{H^+}-\tan\beta$ parameter space is excluded for $m_{H^+}=80-160$~GeV and the only unexcluded regions of this parameter space include the charged Higgs masses as $90\leq m_{H^+}\leq 100$~GeV (for $6<\tan\beta <10$) and  $140\leq m_{H^+}\leq 160$~GeV (for $3<\tan\beta<21$). See also figure 9 of Ref.~\cite{CMS:2014cdp}.
Therefore, in this work our phenomenological predictions  are restricted to these unexcluded regions, however, a definitive search of the charged Higgses over these parts of the $m_{H^+}-\tan\beta$  parameter space  still has to be carried out by the LHC experiments.

In the following, for our numerical analysis we adopt the input parameter values from Ref.~\cite{Olive:2016xmw} as; 
$G_F = 1.16637\times10^{-5}$~GeV$^{-2}$,
$m_t = 172.98$~GeV,
$m_b=4.78$~GeV,
$m_W=80.399$~GeV,
$m_B = 5.279$~GeV, and
$|V_{tb}|=0.999152$, 
and from  the unexcluded $m_{H^+}-\tan\beta$ parameter space determined by  the ATLAS experiments \cite{TheATLAScollaboration:2013wia}, we also adopt $m_{H^+}=95, 155$~GeV and $160$~GeV. \\
Now, we present and compare our results for the NLO decay widths  $\Gamma(t(\uparrow)\to bH^+)$  in the ZM- and GM-VFN schemes in both models.  Considering $m_{H^+}=160$~GeV, one has 
\begin{eqnarray}\label{aval}
	\frac{\tilde\Gamma^{\textbf{pol}}_{\textbf{NLO}}}{\tilde\Gamma_{0P}}&=& 1-0.026\quad \textrm{for Type-I Scenario }
	\nonumber\\
	\frac{\tilde\Gamma^{\textbf{pol}}_{\textbf{NLO}}}{\tilde\Gamma_{0P}}&=& 1-0.591\quad \textrm{for Type-II Scenario}\quad (\tan\beta=10)
	\nonumber\\
	\frac{\tilde\Gamma^{\textbf{pol}}_{\textbf{NLO}}}{\tilde\Gamma_{0P}}&=& 1-0.523\quad \textrm{for Type-II Scenario}\quad (\tan\beta=16)
	\nonumber\\
	\frac{\hat\Gamma^{\textbf{pol}}_{\textbf{NLO}}}{\hat\Gamma_{0P}}&=& 1-0.033
	\quad \textrm{in ZM-VFNs  for both scenarios}
	\nonumber\\
	\end{eqnarray}
and for $m_{H^+}=95$~GeV, they read 
\begin{eqnarray}\label{dovom}
\frac{\tilde\Gamma^{\textbf{pol}}_{\textbf{NLO}}}{\tilde\Gamma_{0P}}&=& 1-0.118\quad \textrm{for Type-I Scenario}
\nonumber\\
\frac{\tilde\Gamma^{\textbf{pol}}_{\textbf{NLO}}}{\tilde\Gamma_{0P}}&=& 1-0.681\quad \textrm{for Type-II Scenario}\quad (\tan\beta=10)
\nonumber\\
\frac{\tilde\Gamma^{\textbf{pol}}_{\textbf{NLO}}}{\tilde\Gamma_{0P}}&=& 1-0.617\quad \textrm{for Type-II Scenario}\quad (\tan\beta=16)
\nonumber\\
\frac{\hat\Gamma^{\textbf{pol}}_{\textbf{NLO}}}{\hat\Gamma_{0P}}&=& 1-0.109
\quad \textrm{in ZM-VFNs for both scenarios}
\nonumber\\
\end{eqnarray}
Note that the normalized decay rates in the ZM-VFNS ($\hat\Gamma^{\textbf{pol}}_{\textbf{NLO}}/\hat{\Gamma}_{0P}$) are independent of the models while in the GM-VFN scheme the normalized widths of the polarized top decays ($\tilde\Gamma^{\textbf{pol}}_{\textbf{NLO}}/\tilde{\Gamma}_{0P}$) depend on the model selected, extremely. Also, the results in the Type-I 2HDM scenario are independent of $\tan\beta$, while the Type-II 2HDM  results depend on the $\tan\beta$.

Here, we are in a situation to present our results for the scaled-energy  distribution of hadrons inclusively produced in polarized top decays in two variants of the 2HDM. Since the bottom quarks produced through the top decays hadronize before they decay and each b-jet contains a bottom-flavored hadron which most of the times is a B-meson, then we study the energy distribution of B-mesons. For this study, we
consider the quantity $d\Gamma(t(\uparrow)\to BH^++X)/dx_B$.
According to  the factorization formula (\ref{azi}), to evaluate $d\Gamma/dx_B$ one needs  the parton-level decay width ($d\Gamma^{\textbf{pol}}_{\textbf{GM}}/dx_b$) described in section~\ref{sec:two}, and the nonperturbative fragmentation function $D_b^B(z,\mu_F)$ which describes the splitting of $b\to B$ at the desired scale $\mu_F$.
To describe the hadronization process $b\to B$, from Ref.~\cite{Kniehl:2008zza} we adopt the nonperturbative fragmentation function $D_b^B(z,\mu_F)$ determined at NLO through a global fit  to
electron-positron annihilation data taken by OPAL
\cite{Abbiendi:2002vt}, ALEPH \cite{Heister:2001jg} and SLD \cite{Abe:1999ki}. In \cite{Kniehl:2008zza}, a simple power model as  $D_b(z,\mu_F^\text{ini})=Nz^\alpha(1-z)^\beta$ is proposed as a initial condition for the $b\to B$ FF at the initial scale $\mu_F^\text{ini}=4.5$~GeV. Their fit results for the FF parameters read: $N=4684.1$, $\alpha=16.87$, and $\beta=2.628$. The nonperturbative FF $D_b^B(z,\mu_F)$ at each desired scale might be generated via the DGLAP evolution equations \cite{dglap}. Note that, in the factorization formula (\ref{azi}) the factorization ($\mu_F$) and renormalization ($\mu_R$)  scales are completely arbitrary and, in principle, one can select different values for them. In this work, we adopt $\mu_R=\mu_F=m_t$ that  allows us to get rid of the term $\ln(\mu_F^2/m_t^2)$ in the the partial decay rate in the ZM-VFNs ($d\hat\Gamma^{\textbf{pol}}_{\textbf{NLO}}/dx_b$), see the analytical result in \cite{MoosaviNejad:2016jpc}. In \cite{MoosaviNejad:2016jpc}, we also investigated the dependence of the B-meson energy
spectrum on these scales considering  two other different values: $\mu_R=\mu_F=m_t/2$ and $\mu_R=\mu_F=2m_t$. Since in  Ref.~\cite{Kniehl:2008zza} no uncertainty is reported for the nonperturbative FF $D_b^B(z,\mu_F)$, then 
this scale variation can be considered as a just theoretical uncertainty.
\begin{figure}
	\begin{center}
		\includegraphics[width=0.7\linewidth,bb=137 42 690 690]{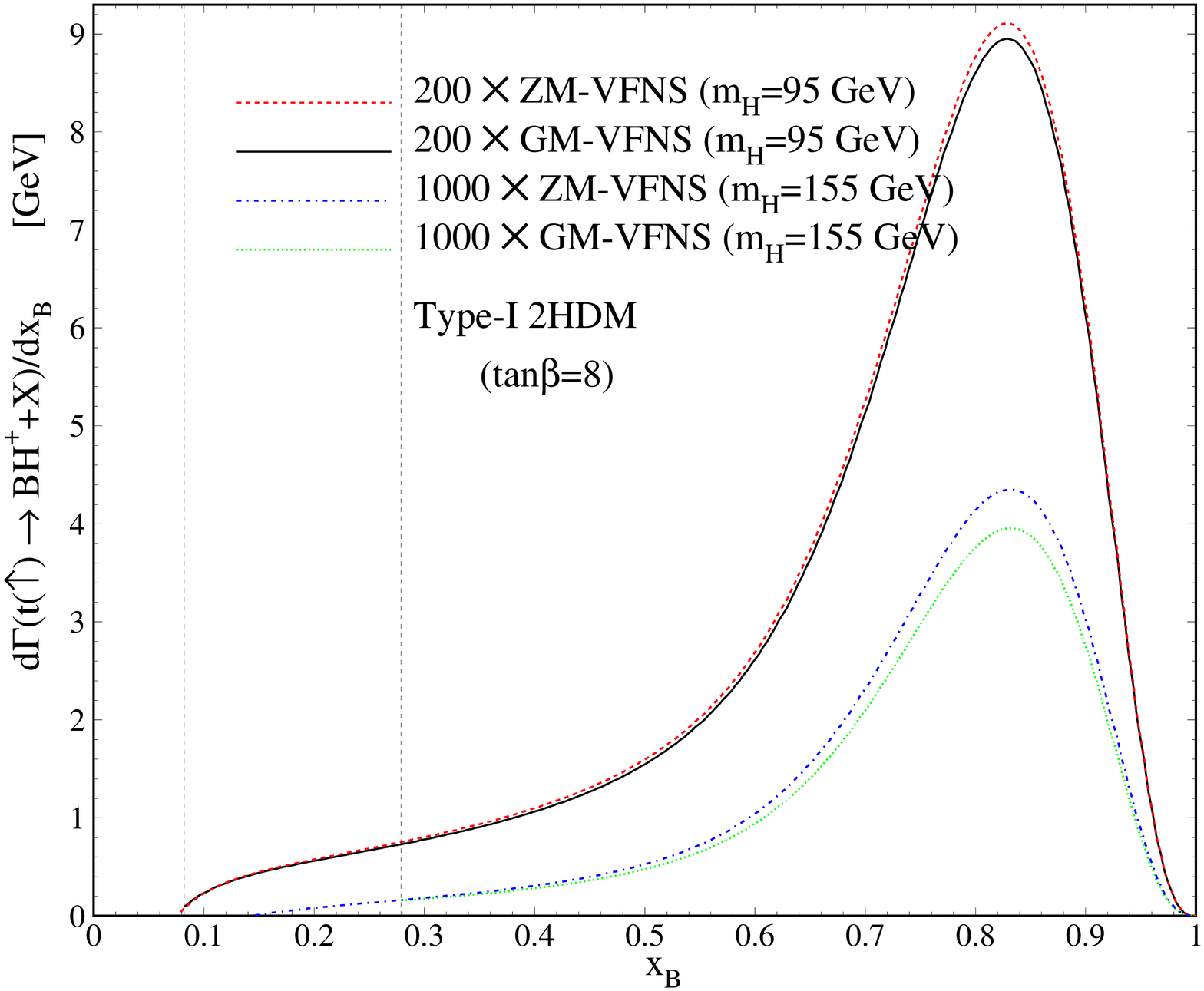}
		\caption{\label{fig3}%
			Comparison of the $x_B$ spectrum in the ZM- and GM-VFN schemes in the Type-I 2HDM scenario at the scale $\mu_F=m_t$. Different values of the Higgs boson mass are considered, i.e. $m_{H^+}=95$~GeV and $m_{H^+}=155$~GeV. Another free parameter is fixed to $\tan\beta=8$. Thresholds are also shown.}
	\end{center}
\end{figure}
\begin{figure}
	\begin{center}
		\includegraphics[width=0.7\linewidth,bb=137 42 690 690]{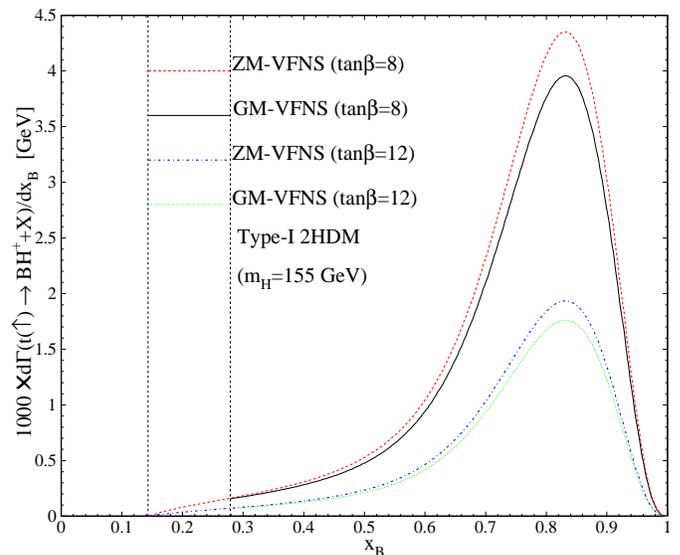}
		\caption{\label{fig4}%
			$x_B$ spectrum  in polarized top decay in the Type-I 2HDM. The GM-FVNS results are compared to the ZM-VFNS ones using $\tan\beta=8$ and $\tan\beta=12$ while the charged Higgs mass is fixed to $m_{H^+}=155$~GeV. Thresholds are also shown.}
	\end{center}
\end{figure}
\begin{figure}
	\begin{center}
		\includegraphics[width=0.7\linewidth,bb=137 42 690 690]{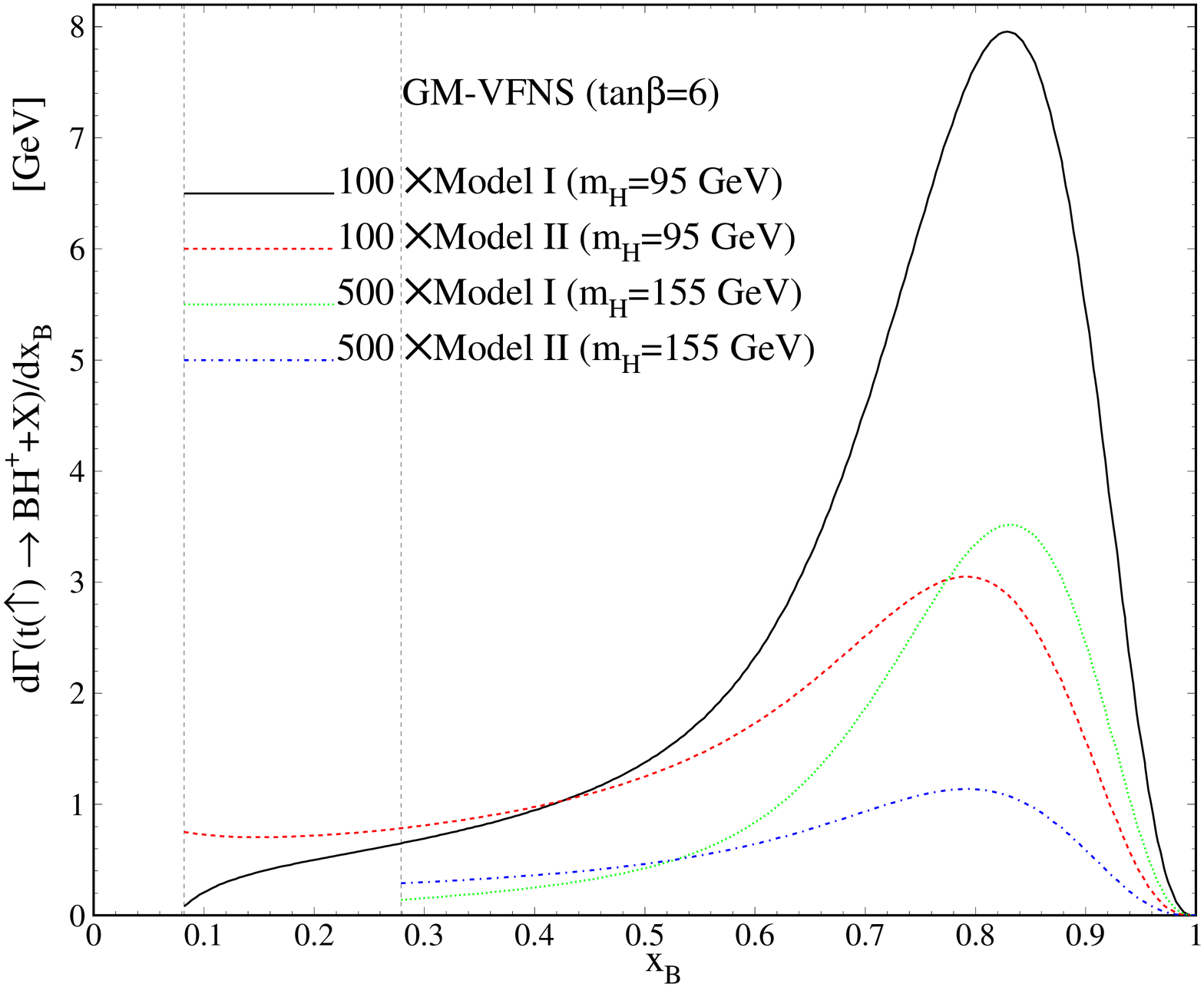}
		\caption{\label{fig5}%
			$d\Gamma/dx_B$ as a function of $x_B$ in the Type-I and Type-II 2HDM scenarios considering the GM-VFN ($m_b\neq 0$) scheme.  We set  $m_{H^+}=95$ and $155$~GeV taking  $\tan\beta=6$. This plot can be also considered as the results for the model II in the ZM- and GM-VFN schemes.} 
	\end{center}
\end{figure}

Considering the unexcluded MSSM $m_{H^+}-\tan\beta$ parameter space from 
the CMS \cite{CMS:2014cdp} and ATLAS \cite{TheATLAScollaboration:2013wia} experiments, in Fig.~\ref{fig3} we present our prediction for the $x_B$-spectrum at NLO taking $\tan\beta=8$.
Our results for the ZM- and GM-VFN schemes are compared in the model I, taking $m_{H^+}=95$~GeV and $m_{H^+}=155$~GeV. As is seen the zero-mass approximation (with $m_b=0$), applied in our previous work \cite{MoosaviNejad:2016jpc}, works good to a high accuracy. Here, the B-hadron mass creates a
threshold, e.g., at $x_B=2m_B/(m_t(1+R-y))\approx 0.28$ for $m_{H^+}=155$~GeV when $m_b\neq 0$.

In Fig.~\ref{fig4},  considering the model I of 2HDM we study the energy spectrum of B-hadron in the ZM- and GM-VFN schemes for 
different values of $\tan\beta=8$ and $12$, where the mass of Higgs boson is fixed to $m_{H^+}=155$~GeV. As is seen, when $\tan\beta$ increases the decay rate  decreases in both schemes, as $\tilde\Gamma_{0P}$  is proportional to $\cot^2\beta$, see (\ref{gammatree}) and (\ref{haselmodel1}). As in Fig.~\ref{fig3}, the results of massless and massive schemes are, with a good approximation, the same so the results of ZM-VFN scheme  show an enhancement in the size of decay rates, specifically at $x_B=0.8$.

In Fig.~\ref{fig5}, considering the GM-VFN scheme the NLO energy spectrum of B-hadrons are compared in both models of the 2HDM (Type-I and II) for $m_{H^+}=95$~GeV and $m_{H^+}=155$~GeV where $\tan\beta$ is fixed to $\tan\beta=6$. As is seen, the B-hadron energy spectrum  in the 2HDM extremely depends on the model and one can not also naively set $m_b=0$ in all expressions in the model II. Also, the results from the model I are always larger than the ones from the model II in most of $x_B$-regions.\\
Note that, in one hand,  the results of the model I do not depend on the b-quark mass largely (Figs.~\ref{fig3} and \ref{fig4}) and, on the other hand, in the limit $m_b\to 0$ the results of both models are the same (see Eqs.~(\ref{haselmodel1}) and (\ref{haselmodel2})), then the results shown in Fig.~\ref{fig5} for the model I (solid and dotted lines), in principle, can be assumed as the results of model II in the ZM-VFN scheme. In fact, the results shown in Fig.~\ref{fig5} can be also considered as the results for the model II in the ZM- and GM-VFN schemes and this figure shows that there is an egregious  difference between the massless and massive schemes.

Finally, the results from Figs.~\ref{fig3}-\ref{fig5}, specially  Fig.~\ref{fig5}, show that the most reliable results are made in the GM-VFN  scheme, specifically, when the Type-II 2HDM scenario is concerned.

\section{Conclusions}
\label{sec:four}

Charged Higgs bosons $H^\pm$ are predicted in many extensions of the Standard Model consist of, at least,  two Higgs doublets, of which the simplest are the two-Higgs-doublet models (2HDM) so the discovery of them would clearly indicate  unambiguous evidence for the presence of new physics beyond the SM.
Although, the charged Higgs bosons have been searched for in high energy experiments, in particular, at the Tevatron, ATLAS and CMS but they have  not been seen so far and a definitive search is a program that still has to be carried out by the CERN LHC.

In this work, which is a fundamental extension of our previous work \cite{{MoosaviNejad:2016jpc}}, we introduce a channel to indirect search for the charged Higgs bosons. In fact, since the main production mode of light charged Higgses in the 2HDM is through the top quark decay, $t\to bH^+$, and whence bottom quarks hadronize ($b\to B$) before they decay, then the study of B-meson energy spectrum in the decay mode $t\to BH^++X$ would be of prime importance at the LHC. 
In other words, at the LHC any deviation of the B-meson energy spectrum  from the SM predictions \cite{Nejad:2016epx} can be considered as a signal for the existence of charged Higgs.

In \cite{MoosaviNejad:2016jpc}, using the massless or ZM-VFN scheme (with $m_b=0$) we studied the spin-dependent energy distribution of B-mesons ($d^2\hat\Gamma^{\textbf{pol}}/(dx_B d\cos\theta_P)$) produced through the polarized top  decays in a special helicity coordinate system, where the event plane lies in the $(\hat{x}, \hat{z})$ plane and the b-quark three-momentum is considered along the $\hat{z}$-axis. In this system the polarization vector of the top quark is evaluated relative to the b-quark three-momentum. 
In the ZM-VFN scheme, since all information on the $m_b$ dependence of the B-hadron spectrum is wasted, then our results are reliable just for the Type-I 2HDM scenario.  
In the present work, we studied the same decay mode in the GM-VFN scheme where $m_b\neq 0$ is considered from the beginning, however, considering the b-quark mass effects makes the calculations so complicated. Unlike the massless results, the massive decay rates  are extremely dependent on the scenario selected in the 2HDM, specifically, when the type-II 2HDM scenario is concerned.
Our results show  that the most reliable predictions for the B-hadron energy spectrum are made in the GM-VFN scheme. 
 
Note that, since for the branching ratio of the decay $H^+\to \tau^+\nu_\tau$ one has $BR(H^+\to \tau^+\nu_\tau)=1$  to a very high accuracy, then the results presented in this work for $d\Gamma(t(\uparrow)\to BH^++X)/dx_B$ are also valid for $d\Gamma(t(\uparrow) \to B\tau^+\nu_\tau+X)/dx_B$.

Our formalism elaborated in this work can be also extended  to  other hadrons, such as pions, kaons and protons, etc.,  using the nonperturbative $b\rightarrow \pi/K/P$ FFs extracted in \cite{Soleymaninia:2013cxa,Nejad:2015fdh},
relying on their universality and scaling violations.

\section{Acknowledgments}
\label{sec7}
We would like to thank  the LHC top working group  for importance discussion and comments. We warmly acknowledge the CERN TH-PH division for its hospitality where a portion of this work was performed.


\begin{thebibliography}{32}
	
	\bibitem{Inoue:1982pi}
	K.~Inoue, A.~Kakuto, H.~Komatsu and S.~Takeshita,
	Prog.\ Theor.\ Phys.\  {\bf 68} (1982) 927;
	Erratum: [Prog.\ Theor.\ Phys.\  {\bf 70} (1983) 330].
	
	
	\bibitem{Lee:1973iz}
	T.~D.~Lee,
	Phys.\ Rev.\ D {\bf 8} (1973) 1226.
	

	\bibitem{Gunion}
	J.~F.~Gunion and H.~E.~Haber,
	Nucl.\ Phys.\  B {\bf 272}, 1  (1986);  {\bf 402}, 567 (1993).
	
	\bibitem{Djouadi:2005gj}
	A.~Djouadi,
	Phys.\ Rept.\  {\bf 459}, 1 (2008)  [hep-ph/0503173].  
		
	\bibitem{Cabibbo:1963yz}
	N.~Cabibbo,
	Phys.\ Rev.\ Lett.\  {\bf 10}, 531 (1963);
	M.~Kobayashi and T.~Maskawa,
	Prog.\ Theor.\ Phys.\  {\bf 49}, 652 (1973).
		
	\bibitem{Langenfeld:2010zz}
	U.~Langenfeld, S.~Moch, P.~Uwer, arXiv:0907.2527 [hep-ph].
	

	\bibitem{CMS:2014cdp}
	CMS Collaboration [CMS Collaboration],
	CMS-PAS-HIG-14-020;
	V.~Khachatryan {\it et al.} [CMS Collaboration],
	JHEP {\bf 1511} (2015) 018.
	
	
	
	\bibitem{TheATLAScollaboration:2013wia}
	The ATLAS collaboration [ATLAS Collaboration],
	ATLAS-CONF-2013-090.
	
	
	
	\bibitem{collins}
	J.~C.~Collins,
	Phys.\ Rev.\  D {\bf 58}, 094002 (1998).
	
	
	\bibitem{MoosaviNejad:2011yp}
	S.~M.~Moosavi Nejad,
	Phys.\ Rev.\ D {\bf 85}, 054010 (2012);
	Eur.\ Phys.\ J.\ C {\bf 72} (2012) 2224.
	
	\bibitem{MoosaviNejad:2016jpc}
	S.~M.~Moosavi Nejad and S.~Abbaspour,
	JHEP {\bf 1703} (2017) 051.
	
	
	\bibitem{Binnewies:1998vm} 
	J.~Binnewies, B.~A.~Kniehl and G.~Kramer,
	Phys.\ Rev.\ D {\bf 58}, 034016 (1998).
	
	\bibitem{Kniehl:2012mn}
	B.~A.~Kniehl, G.~Kramer and S.~M.~M.~Nejad,
	Nucl.\ Phys.\  B {\bf 862}, 720  (2012).
	
	
	\bibitem{Nejad:2015pca}
	S.~M.~Moosavi Nejad,
	Nucl.\ Phys.\ B {\bf 905} (2016) 217.
	
	
	
	\bibitem{Nejad:2016epx}
	S.~M.~Moosavi Nejad and M.~Balali,
	Eur.\ Phys.\ J.\ C {\bf 76} (2016) no.3,  173.
	
	\bibitem{Nejad:2013fba}
	S.~M.~Moosavi Nejad,
	Phys.\ Rev.\ D {\bf 88} (2013) no.9,  094011.
	
	\bibitem{Nejad:2014sla}
	S.~M.~Moosavi Nejad and M.~Balali,
	Phys.\ Rev.\ D {\bf 90} (2014) no.11,  114017.
	
	
	\bibitem{Ali:2011qf}
	A.~Ali, F.~Barreiro and J.~Llorente, Eur.\ Phys.\ J.\  C {\bf 71}, 1737 (2011).
	
	\bibitem{Ali:2009sm}
	A.~Ali, E.~A.~Kuraev and Y.~M.~Bystritskiy,
	Eur.\ Phys.\ J.\ C {\bf 67} (2010) 377.
	
			
	\bibitem{GHK}
	J.~F.~Gunion, H.~Haber, G.~Kane, and S.~Dawson,\textit{ The Higgs Hunter's Guide}
	(Addison-Wesley, Reading, MAA, 1990), and refrences therein.
	
		
		
	\bibitem{Barger:1989fj}
	V.~D.~Barger, J.~L.~Hewett and R.~J.~N.~Phillips,
	Phys.\ Rev.\ D {\bf 41} (1990) 3421.
		
	\bibitem{kadeer}
	A.~Kadeer, J.~G.~K\"orner, and M.~C.~Mauser,
	Eur.\ Phys.\ J.\  C {\bf 54}, 175 (2008).	
		
	\bibitem{Liud}
	J.~Liu and Y.~P.~Yao,
	Phys.\ Rev.\  D {\bf 46}, 5196 (1992).
	
	\bibitem{Czarnecki}
	A.~Czarnecki and S.~Davidson,
	Phys.\ Rev.\  D {\bf 47}, 3063 (1993).
	
	
	\bibitem{Dittmaier:2003bc}
	S.~Dittmaier,
	Nucl.\ Phys.\ B {\bf 675}, 447 (2003).  
	
	
	\bibitem{Corcella:1}
	G.~Corcella and A.~D.~Mitov,
	Nucl.\ Phys.\  B {\bf 623}, 247 (2002).
	
	
	\bibitem{MoosaviNejad:2016aad}
	S.~M.~Moosavi Nejad and S.~Abbaspour,
	arXiv:1610.03811 [hep-ph].
	
	
		\bibitem{Kniehl:2}
		B.~A.~Kniehl, G.~Kramer, I.~Schienbein and H.~Spiesberger,
		Phys.\ Rev.\ D {\bf 71}, 014018 (2005).  
		\bibitem{Kniehl:3}
		B.~A.~Kniehl, G.~Kramer, I.~Schienbein and H.~Spiesberger,
		Phys.\ Rev.\ Lett.\  {\bf 96}, 012001 (2006).  
	
		\bibitem{Mele:1990cw}
		B.~Mele, P.~Nason,
		Nucl.\ Phys.\ B 361 (1991) 626;\\
		J.P.~Ma,
		Nucl.\ Phys.\ B 506 (1997) 329;\\
		S.~Keller, E.~Laenen,
		Phys.\ Rev.\ D 59 (1999) 114004;\\
		M.~Cacciari, S.~Catani,
		Nucl.\ Phys.\ B 617 (2001) 253;\\
		K.~Melnikov, A.~Mitov,
		Phys.\ Rev.\ D 70 (2004) 034027;\\
		A.~Mitov,
		Phys.\ Rev.\ D 71 (2005) 054021.
	
	\bibitem{Nakamura:2010zzi}
	K.~Nakamura {\it et al.}\  (Particle Data Group),
	J.\ Phys.\ G {\bf 37}, 075021 (2010).
	
	\bibitem{Olive:2016xmw}
	C.~Patrignani {\it et al.} [Particle Data Group],
	Chin.\ Phys.\ C {\bf 40} (2016) no.10,  100001.
	
	
	
		\bibitem{Kniehl:2008zza}
		B.~A.~Kniehl, G.~Kramer, I.~Schienbein, and H.~Spiesberger,
		Phys.\ Rev.\  D {\bf 77}, 014011 (2008).
		
		\bibitem{Abbiendi:2002vt}
		G.~Abbiendi {\it et al.}\  (OPAL Collaboration),
		Eur.\ Phys.\ J.\  C {\bf 29}, 463 (2003).
		
		\bibitem{Heister:2001jg}
		A.~Heister {\it et al.}\  (ALEPH Collaboration),
		Phys.\ Lett.\  B {\bf 512}, 30 (2001).
		
		\bibitem{Abe:1999ki}
		K.~Abe {\it et al.}\  (SLD Collaboration),
		Phys.\ Rev.\ Lett.\  {\bf 84}, 4300 (2000);
		Phys.\ Rev.\  D {\bf 65}, 092006 (2002);
		{\bf 66}, 079905 (2002).
		
		\bibitem{dglap}
		V.~N.~Gribov and L.~N.~Lipatov,
		Sov.\ J.\ Nucl.\ Phys.\  {\bf 15}, 438 (1972)
		[Yad.\ Fiz.\  {\bf 15}, 781 (1972)];
		G.~Altarelli and G.~Parisi,
		Nucl.\ Phys.\ {\bf B126}, 298 (1977);
		Yu.~L.~Dokshitzer,
		Sov.\ Phys.\ JETP {\bf 46}, 641 (1977)
		[Zh.\ Eksp.\ Teor.\ Fiz.\  {\bf 73}, 1216 (1977)].
		
	\bibitem{Soleymaninia:2013cxa}
	M.~Soleymaninia, A.~N.~Khorramian, S.~M.~Moosavi Nejad and F.~Arbabifar,
	Phys.\ Rev.\ D {\bf 88} (2013) no.5,  054019.
	
	\bibitem{Nejad:2015fdh}
	S.~M.~Moosavi Nejad, M.~Soleymaninia and A.~Maktoubian,
	Eur.\ Phys.\ J.\ A {\bf 52} (2016) no.10,  316.
		
	
	
\end{thebibliography}
\end{document}